\documentclass[10pt, conference]{IEEEtran}
\IEEEoverridecommandlockouts
\usepackage{cite}
\usepackage{amsmath,amssymb,amsfonts}
\usepackage[linesnumbered, ruled,vlined]{algorithm2e}
\usepackage{algorithmic}
\usepackage{graphicx}
\usepackage{textcomp}
\usepackage{xcolor}
\usepackage{multirow}
\usepackage{color}

\def\BibTeX{{\rm B\kern-.05em{\sc i\kern-.025em b}\kern-.08em
    T\kern-.1667em\lower.7ex\hbox{E}\kern-.125emX}}
\begin{document}

\title{Feedback-based, Automated Failure Testing of Microservice-based Applications}

\author{\IEEEauthorblockN{Chengxu Cui, Guoquan Wu, Wei Chen, Jiaxing Zhu, Jun Wei}
\IEEEauthorblockA{State key Lab, Institute of Software, Chinese Academy of Sciences \\ University of Chinese Academy of Sciences,Beijing, China \\ 
Email: {\{cuichengxu16, gqwu,wchen, zhujiaxin, wj\}}@otcaix.iscas.ac.cn}}
\maketitle

\begin{abstract}
Modern distributed applications are moving toward a microservice architecture, in which each service is developed and managed independently, and new features and updates are delivered continuously. A guiding principle of microservice architecture is that it must be built to anticipate and mitigate a variety of hardware and software failures. In order to test the fault handling capabilities of microservces, this paper presents IntelliFT, a feedback-based, automated failure testing technique for microservice based applications, which aims to expose the defects in the fault-handling logic quickly. The initial experimental result on a medium-size microservice benchmark system shows that the proposed approach is effective.
\end{abstract}


\section{Introduction}

Microservice architecture \cite{newman2015building} is gaining more and more popularity in the industry when designing complex, cloud-based distributed systems. In this architecture, the application consists of small, loosely coupled, mono-functional services that communicate using REST-like interfaces over the network. Each microservice is developed, deployed and managed independently; new features and updates are delivered continuously, resulting in polyglot applications that are extremely dynamic and rapidly evolved. Many organizations, like Netflix, Amazon, eBay and Twitter, have already evolved their applications to microservice architecture.

In order to provide an ``always on" experience to customers, microservice should be designed to anticipate, detect and withstand various runtime failures and outages from its dependencies, and struggle to remain available when deployed. However, it is difficult to ensure that such fault-tolerant code is adequately tested. In the past, many popular highly available Internet service have experienced various failures and outages (e.g., cascading failures due to database overload). The post-mortem reports revealed that such outages were mainly caused by missing or faulty failure handling logic, with an acknowledgement that unit and integration testing are insufficient to catch bugs in the failure recovery logic \cite{heorhiadi2016gremlin}.

Chaos engineering \cite{basiri2016chaos}, the practice of performing fault injection experiments on the production system, to increase the overall resilience of a software system, is emerging as a discipline to tackle the resilience of large-scale distributed system. It was pioneered by Netflix, which developed Chaos Monkey \cite{ChaosMonkey} to inject faults randomly in production system to test the resiliency of the AWS. Since then, Linkedin, Microsoft and Uber have developed fault injection framework to improve the resilience of their systems at scale. 

However, random fault injection technique is not efficient, and lots of time and resources can be wasted to explore redundant failure scenarios. It is also unlikely to uncover deep failures involving combinations of different instances and kinds of faults. To address these issues,  Alvaro et al. \cite{alvaro2015lineage} proposed lineage-driven fault injection (LDFI) technique, to discover bugs in fault tolerant protocols/systems. It combines data lineage from database literature and satisfiability testing to infer backwards from correct system outcomes to determine whether injected faults in the execution could prevent the outcome. They also adapted LDFI and implemented a research prototype to automate failure testing of Netflix microservice platform \cite{alvaro2016automating}. 

Although LDFI technique has achieved promising results for testing the resilience of microservice based applications, there are still room for improvement. Firstly, it focuses on the systematic exploration of the fault space for each user request independently without leveraging the historical testing results to optimize fault injections when exploring the fault space that other user requests expose. Secondly, although LDFI technique computes the minimized fault injection points for individual user request, the solutions are based on the current observed execution traces. In fact, some injection points do not need to be explored based on the observed testing results of previous fault injections (we will explain this in Section III). 

To address above limitations, in this paper, we present IntelliFT, a feedback-based, automated failure testing technique to expose fault-tolerance bugs in microservice based applications quickly. Firstly, leveraging the historical fault injection results for different user requests, we design a novel feedback-based fault space exploration algorithm. Guided by the dynamically updated priority values, it can optimize the selection of fault injection points to expose fault tolerance bugs more effectively. Secondly, by observing the propagation of injected faults and testing results, we further propose two heuristic rules, which can reduce unnecessary fault injections computed by LFDI technique. Our initial experimental result shows that the proposed approach is effective. In summary, our contribution are as follows:

\begin{itemize}
  \item We design a novel feedback-based, automated failure testing technique for microservice based applications, which leverages the historical fault injection results to guide the fault space exploration effectively across different user requests;
  \item We propose two heuristic rules which can avoid unnecessary fault injections by observing the propagation of injected faults and testing results; 
  \item Leveraging existing service mesh framework ---Istio \cite{Istio}), we implemented an initial prototype to execute failure testing automatically.
\end{itemize}

\section{Background and Motivation}

Lineage-driven fault injection (LDFI) is a technique which leverages data lineage in the database community and satisfiability testing to reason backwards from correct system outcomes to determine whether some combinations of faults could have prevented the outcome. LDFI is based on two key insights. The first is that fault-tolerance is redundancy, and a fault-tolerant system/program can provide alternative way to obtain an expected outcome in the presence of some common faults (e.g., component failure). The second insight is that instead of exhaustively exploring the space of all possible executions from initial state, a better strategy to quickly expose fault-tolerance bugs is to start with successful outcomes and reason backwards, to understand whether some combination of faults could prevent the outcome.

Generally, LDFI technique works as follows: firstly, the system under test is evaluated by performing a failure-free execution. Then by analyzing why successful outcome is achieved, lineage graph can be extracted. It is further converted into CNF formula that is passed to a SAT solver to generate failure hypothesis. The solved combination of faults will be transformed into inputs that try to falsify the successful execution in the next round of the loop. This process continues until either a fault tolerance bug is identified or the system exhausts its resources. 

Netflix also adapted this technique to enable automated failure testing of microservice based applications \cite{alvaro2016automating}. We explain how it works using a concrete example. Figure 1(a) shows a failure-free service call graphs for a user request. The execution path is first transformed into formula: $API \vee Review \vee Rating \vee PlayList$. The solution to this boolean represents the sets of faults that we should test via fault injection. Using SAT, the minimal solutions based on current observed execution are: \{$API$\}, \{$Review$\}, \{$Rating$\} and \{$PlayList$\}. These solved sets are referred to as \textit{injection points} in this paper, and can be used to create different \textit{failure scenarios} --- the sets of injection points into which different fault (e.g., abort, message delay) can be injected. In the next loop, LDFI chooses an injection point set (e.g., \{$Ratings$\}) from the current solutions, and inject a fault. This time, after sending the request, the user still gets the successful response. Figure 1(b) shows the corresponding service call graphs after injecting a fault in \textit{Rating} service. It can be seen that when \textit{Rating} service fails, a backup path ($Review$ service calls $Rating_{cache}$) will appear to provide high reliability. The execution path is encoded as $API \vee Review \vee Rating_{cache} \vee PlayList$, which is conjuncted with the previous formula: $API \vee Review \vee Rating \vee PlayList$. Using SAT solver, the minimal solutions to invalid both execution paths are: \{$API$\}, \{$Review$\}, \{$Rating$,$Rating_{cache}$\} and \{$PlayList$\}. Based on the solved solutions, LDFI will continue to explore different failure scenarios to find the defects in the fault handling logic.

\begin{figure}[hbt]
  \includegraphics[width=0.45\textwidth]{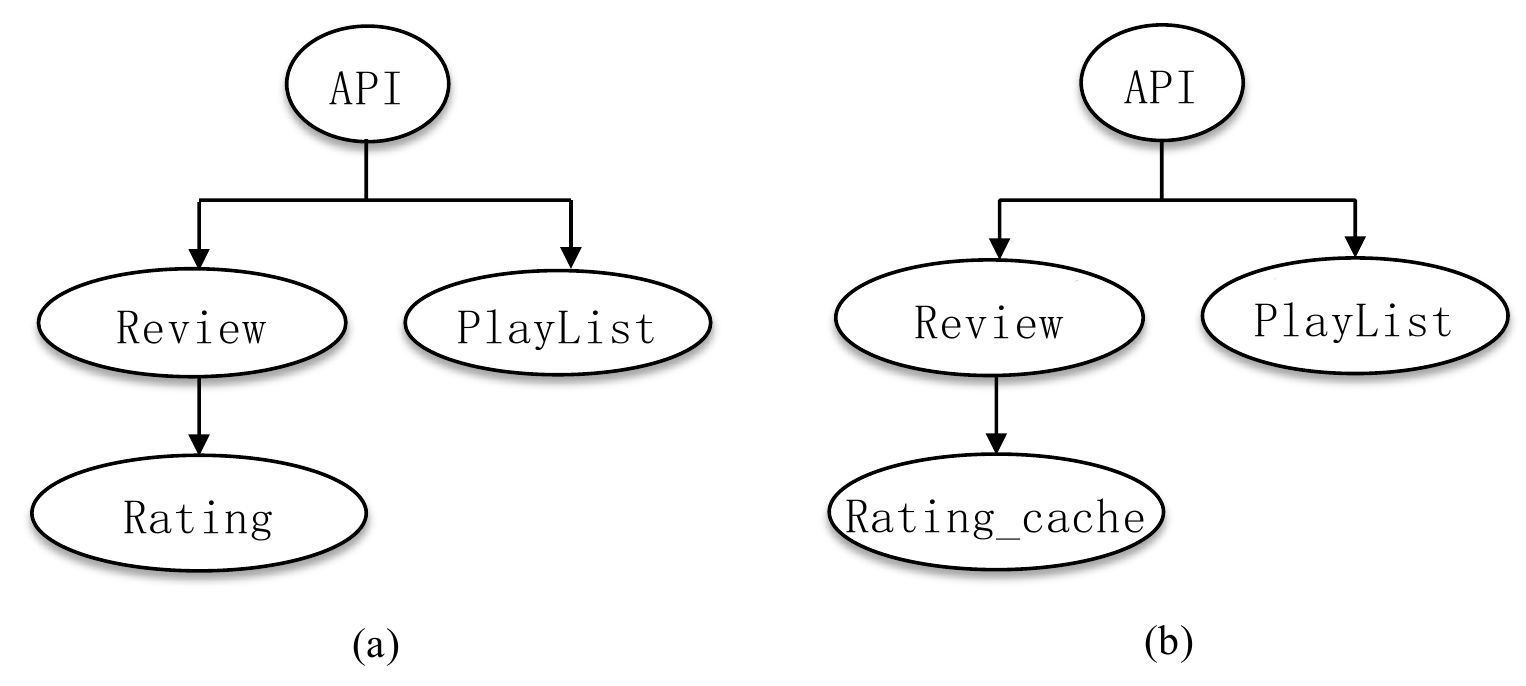}
  \caption{service call graph without/with fault injection}
  \vspace{-10px}
\end{figure}

Compared with random strategy, LDFI technique can expose some ``deep" bugs involving complex failure scenarios (e.g., multiple injection points and fault types). Although LDFI technique has achieved promising result in Netflix, some problems are still not addressed. For example, LDFI systematically explores the fault space for each user request independently, and each candidate failure scenario is also treated equally. It does not support to share the historical failure testing results across different user requests to guide the selection of failure scenarios that will have high impact on the reliable execution of the application. 

In practice, there will be lots of different user requests for complex microservice based applications, and each request may touch dozens of internal services and hundreds of potential injection points. Therefore, a large number of fault space needs to be explored. Given limited testing resources/time, it is important if we can optimize fault space exploration and test the most likely faults which can cause severe failure (e.g., user-visible error).

\section{The overall approach}

To improve the state-of-the-art of failure testing of microservice based applications, we propose a novel feedback-based fault space exploration technique, which leverages historical fault injection results to optimize the fault space exploration across different user requests. Moreover, it supports to avoid unnecessary fault injections based on the failure testing results. Generally, it consists of the following two steps: 

\begin{itemize}
  \item Firstly, our approach leverages existing integration test cases to generate the sequences of user requests. Leveraging distributed tracing technique (e.g., Zipkin), it also captures the internal service call graph for each request, which is then encoded as CNF formula to compute the initial sets of injection points by SAT solver.
  \item Secondly, a feedback-based fault space exploration technique is proposed to effectively expose fault-tolerance bugs in the application. It first randomly generates some failure scenarios, evaluates the impact on the testing results. Then, guided by the designed multiple priority functions, the algorithm tends to select the most likely failure scenarios that can have high impact on the execution of the application.
\end{itemize}
In the following, we describe the two steps in detail.

\subsection{Pre-processing}

To generate the sequences of user requests to interact with the application, during the execution of the test cases, IntelliFT deploys a proxy at the client side to capture user request/response. It also records the timestamp of each user action, user request/response and test assertion. At the sever side, IntelliFT will trace the internal service invocations leveraging distributed tracing technique.

Based on the logged information from the client and server, IntelliFT correlates the user request/response with the service call graph. It also transforms the assertion to UI elements into the assertion to response message.

To avoid repeatedly fault injections for the same type of user requests, it groups different user requests into the same class (similar to \cite{alvaro2016automating}) if they trigger the same set of service invocations at the sever side. IntelliFT then computes the sets of injection points for each request class (using SAT solver) based on the observed service execution path. 

\subsection{Feedback-based Fault Space Exploration}

\subsubsection{Priority function}
 To expose fault-tolerance bugs more effectively, we designed a novel feedback-based fault space exploration technique, to select the most likely failure scenarios across different user requests that can have high impact on the application's execution. The proposed selection strategy is based on three priority functions, which computes the priority of different request classes, the priority of different fault types for each request class and the priority of different microservices, respectively. After testing a failure scenario, the corresponding priority value will be dynamically updated to guide the selection of next failure scenario.
 
 The priority function of each user request class is defined as follows: 
 $$priority(req) = \frac{w_1*num(req.FS)+w_2*num(req.error)}{w_3*num(req.history)}$$

Here, $req.FS$ denotes failure scenarios to be tested, $req.history$ denotes the tested failure scenarios and $req.error$ denotes the exposed unique failures currently. The meaning of this function is that the more fault space (to be explored) and unique failures exposed, the more likely the request will be selected during the failure testing, but with the increase of tested failure scenarios, the opportunity that the request is selected will decrease ($w_1,w_2, w_3$ are the weight parameters).

Each request class also maintains a priority value for each fault type, which is defined as $fail(fault)/num(fault)$, where $fail(fault)$ represents the number of unique failures discovered after injecting the $fault$ and $num(fault)$ represents the number of the injections for this $fault$.

In order to guide the selection of injection points from the candidate solutions, IntelliFT also maintains a priority value for each microservice, which represents the capability of the service to handle the faults. It is defined as $fail(faults,APIs)/sum(faults, APIs)$, where $fail(faults)$ represents the sum of observed failures for the injected faults in the APIs of the service, and $sum(faults, APIs)$ represents the sum of all injected faults in the APIs of the service. The hypothesis behind this definition is that each microservice is developed by individual team, and if one API cannot handle the injected fault, it can be inferred that other APIs of this service may not handle this type of fault with high probability.

\subsubsection{Reduction Rules}
Similar to LDFI, IntelliFT also encodes observed execution paths into CNF formula, and then computes sets of injection points using SAT solver. It then selects a failure scenario based on above defined priority functions. After injecting the fault, based on the observed fault propagation (along the call graph) and the fault injection result, we further define two heuristic rules to reduce unnecessary fault injections in the solutions.

\textbf{(Rule 1)}. If injected fault cannot be handled by its upstreaming services, that is, the same error message is observed to propagate backwards along the call graph, it can be inferred that these upstreaming service calls cannot handle injected faults. IntelliFT can safely avoid to inject the same fault at these upstreaming calls. For instance, consider the call graph in section II, after returning 503 error for the call to $Ratings$ service, if the same error message is observed in the response of $Review$ and $API$ calls, it means both $Review$ and $API$ services cannot handle such fault, and be removed from $req.FS$ set. 

\textbf{(Rule 2)}. For a solution $ip$ in the solved sets of injection points $req.IPS$, if a failure is observed in the user response after injecting fault at $ip$, all other injection points in $req.IPS$ that contain this $ip$ will not inject the same fault. This rule assures that if simple failure scenario cannot be handled successfully by the application, more complex failure scenarios do not need to be tested as the application will be destined to fail. Note that, although the solution in one loop cannot contain each other, the one computed in the subsequent loops may contain the former solution.

\subsubsection{Failure Detection} To detect whether the injected fault is successfully handled by the application. Currently, IntelliFT evaluates the returned user response from the following aspects. It first checks http response code, and it is not 200 (e.g., 503, 404), a failure is detected. For the status code 200, it further checks whether the message content contains some key words, such as ``error" and "exception", and if it is true, a failure is observed. IntelliFT also checks whether the assertion (if it exists) to the response message is violated. Besides functional failures, IntelliFT also concerns performance failure and it sets the upper bound of user request latency (e.g., 2s). During the failure testing, if user request latency exceeds the upper bound, the execution is treated as a failure.

In addition to the above general rules, developers can also define some application specific metrics. By comparing the metric of the application with/with fault injection, a failure can be detected.

\subsubsection{Exploration Algorithm} We describe the proposed exploration algorithm in detail, which mainly consists of two stages: random fault injection (line 2-7) and feedback-based fault space exploration (line 8-13). Firstly, for each request class $req$, function $randSelectFS$ randomly selects a $tc$ (which contains a user request belonged to $req$), an injection point $ip$ and a fault $ft$. Based on the fault injection result (saved by $F_{path}$), function $compPriority$ computes the initial priorities. Then guided by the defined priority functions, function $biasSelectFS$ selects user request, fault type and injection point to expose fault tolerance bugs until the testing resource is exhausted. After injecting the fault in two stages, function $reductFS$ will applies two presented heuristic rules to reduce the generation of unnecessary failure scenarios. 

\begin{algorithm}
\scriptsize
\caption {Feedback-based Fault Space Exploration}
\KwIn{$Class$: maintain the set of request classes;\\
	  $P_{class}$: vector which saves the priority of request classes;\\
	  $P_{req}^{FT}$: vector which saves the priority of faults in $req$;\\
	  $P_{sev}$: vector which saves the priority of services;\\
	  $tc$: the sequence of user requests}
\KwOut{$crash$: saves the observed unique failures}
\Begin {
\ForEach{$req \in Class$}{
	$(tc, ip, ft) \leftarrow randSelectFS()$\;
	$F_{path} \leftarrow injectFault(tc, req, ip, fault)$\;
	$reductFS(F_{path}, req.FS)$\;
	$compPriority(F_{path}, P_{class}, P_{req}^{FT}, P_{services})$\;
	$crash.add(tc, req, ip, F_{path})$\;
	
}
\While {resource is not exhausted}{
    $(tc, req, ft, ip) \leftarrow biasSelectFS(P_{req},req.P_{faults},P_{sev})$\;
    $F_{path} \leftarrow InjectFault(tc, req, fault, ip)$\;
    $reductFS(F_{path}, req.FS)$\;
    $updatePriority(F_{path}, P_{class}, P_{req}^{FT}, P_{sev})$\; 
    $crash.add(tc, req, ip, F_{path})$\;  
}
}
\end{algorithm}

\section{Initial Implementation and Case Study}
The initial implementation of IntelliFT can be found in \textit{https://github.com/ccx1024cc/IntelliFT}. It now supports to explore each user request individually. It leverages the fault injection capability that Isitio provides to simulate typical faults in the microservice based applications. Currently, besides the basic \textit{delay} and \textit{abort} fault, it also supports to simulate \textit{hang}, \textit{disconnect}, \textit{overload}, \textit{abrupt crash} and \textit{transient crash} by composing different scales of \textit{abort} and \textit{delay} faults. For example, \textit{transient crash} is implemented as 10\% \textit{abort} fault with http error code 503. It uses Zipkin to build service call graph, and Z3 solver to compute the minimal sets of injection points. To generate certain amount of user requests, it uses Locust (https://locust.io/) to simulate concurrent user requests. 

Based on the initial prototype, we evaluate the effectiveness of the proposed heuristics rules. The experiment is performed on microservice benchmark ---Trainticket \cite{zhou2018poster} application. As the microservices in this application lack fault handling capabilities for injected faults, we randomly select some service invocations and add specific recovery logic. The injected fault types include \textit{abort} and \textit{delay}. The experiment was done 10 times, and we calculated the average number of fault injections. The result shows that using native LDFI technique, the tool needs to inject faults for 321 times. By applying two reduction rules, the number of fault injections will be decreased to 39 times, showing that the reduction strategy is effective to reduce the exploration of fault space. Moreover, rule 1 has more obvious effect than rule 2 for this application. By analyzing the collected traces, we found that the user requests will trigger many long execution traces, in which only few service invocations can handle the injected faults. Therefore, injected fault will be propagated backwards along the execution path.

\section{Related Work}
Failure testing aims to expose the flaws in fault handling codes, helping developers understand whether applications could survive from unexpected faults. Many approaches have been proposed. Gremlin \cite{heorhiadi2016gremlin} requires developers to design test scripts to test the failure handling capabilities of microservices. Chaos Monkey\cite{ChaosMonkey} is a randomized fault-injection tool, which injects faults by crashing the nodes of specified services. LDFI technique can generate the minimal injection point sets based on observed successful execution. There are also some work which aims to debug microservice based applications. Peng et al. \cite{zhou2018delta} apply delta debugging to minimize failure inducing of circumstances (e.g., environmental configurations). In the work \cite{zhou2018fault}, they further leverages ShiViz \cite{beschastnikh2016debugging}, a debugging visualization tool for distributed system to identify faulty microservice invocation.

\section{Conclusion}
This paper proposes IntelliFT, a feedback-based, automated failure testing technique for microservice based applications, which can optimize the fault space exploration across different user requests. In the future work, we will continue to implement our failure testing tool. We will also conduct more comprehensive studies to evaluate the effectiveness of our approach using real microservice based applications. 

\bibliographystyle{abbrv}
\bibliography{microservice} 

\begin{thebibliography}{10}

\bibitem{Istio}
Istio. retrieved sep. 2018 from https://istio.io/.

\bibitem{ChaosMonkey}
Netflix - chaos monkey released into the wild.
  http://techblog.netflix.com/2012/07/chaos-monkey-released-into- wild.html.

\bibitem{alvaro2016automating}
P.~Alvaro, K.~Andrus, C.~Sanden, C.~Rosenthal, A.~Basiri, and L.~Hochstein.
\newblock Automating failure testing research at internet scale.
\newblock In {\em Proceedings of the Seventh ACM Symposium on Cloud Computing},
  pages 17--28. ACM, 2016.

\bibitem{alvaro2015lineage}
P.~Alvaro, J.~Rosen, and J.~M. Hellerstein.
\newblock Lineage-driven fault injection.
\newblock In {\em Proceedings of the 2015 ACM SIGMOD International Conference
  on Management of Data}, pages 331--346. ACM, 2015.

\bibitem{basiri2016chaos}
A.~Basiri, N.~Behnam, R.~De~Rooij, L.~Hochstein, L.~Kosewski, J.~Reynolds, and
  C.~Rosenthal.
\newblock Chaos engineering.
\newblock {\em IEEE Software}, 33(3):35--41, 2016.

\bibitem{beschastnikh2016debugging}
I.~Beschastnikh, P.~Wang, Y.~Brun, and M.~D. Ernst.
\newblock Debugging distributed systems.
\newblock {\em Communications of the ACM}, 59(8):32--37, 2016.

\bibitem{heorhiadi2016gremlin}
V.~Heorhiadi, S.~Rajagopalan, H.~Jamjoom, M.~K. Reiter, and V.~Sekar.
\newblock Gremlin: Systematic resilience testing of microservices.
\newblock In {\em 2016 IEEE 36th International Conference on Distributed
  Computing Systems (ICDCS)}, pages 57--66. IEEE, 2016.

\bibitem{newman2015building}
S.~Newman.
\newblock {\em Building microservices: designing fine-grained systems}.
\newblock " O'Reilly Media, Inc.", 2015.

\bibitem{zhou2018fault}
X.~Zhou, X.~Peng, T.~Xie, J.~Sun, C.~Ji, W.~Li, and D.~Ding.
\newblock Fault analysis and debugging of microservice systems: Industrial
  survey, benchmark system, and empirical study.
\newblock {\em IEEE Transactions on Software Engineering}, 2018.

\bibitem{zhou2018delta}
X.~Zhou, X.~Peng, T.~Xie, J.~Sun, W.~Li, C.~Ji, and D.~Ding.
\newblock Delta debugging microservice systems.
\newblock In {\em Proceedings of the 33rd ACM/IEEE International Conference on
  Automated Software Engineering}, pages 802--807. ACM, 2018.

\bibitem{zhou2018poster}
X.~Zhou, X.~Peng, T.~Xie, J.~Sun, C.~Xu, C.~Ji, and W.~Zhao.
\newblock Benchmarking microservice systems for software engineering research.
\newblock In {\em IEEE/ACM 40th International Conference on Software
  Engineering: Companion (ICSE-Companion)}, pages 323--324, 2018.

\end{thebibliography}

\end{document}